\documentstyle[aps,pre,preprint]{revtex}

\begin{document}
\author{Guillermo Abramson
\footnote{E-mail: abramson@ictp.trieste.it}
}
\address{International Centre for Theoretical Physics,\\
PO Box 586, 34100 Trieste, Italy}
\title{Ecological model of extinctions}
\date{\today }
\draft
\maketitle

\begin{abstract}
We present numerical results based on a simplified ecological system in
evolution,
showing features of extinction similar to that claimed for the biosystem
on Earth. In the model each species consists of a population in interaction
with the others, that reproduces and evolves in time. Each species is
simultaneously a predator and a prey in a food chain. Mutations that change
the interactions are supposed to occur randomly at a low rate. Extinctions
of populations result naturally from the predator-prey dynamics. The model is not
pinned in a fitness variable, and natural selection arises from the dynamics.
\end{abstract}


\section{Introduction}

The evolution of living organisms is a fascinating phenomenon that has
intrigued the imagination of the scientific and non-scientific community.
However, the formulation of mathematical models falls necessarily to drastic
simplifications. For example, evolution has often been considered as a
``walk'' in a rugged landscape. Following this line, Bak and Sneppen (BS)
have proposed a model of biological evolution \cite{bs} that has become
quite interesting to the physics community due to its simplicity and the new
insight it provides to the problem. It has been shown that this model
evolves to a self-organized critical state (SOC), and is kept there by the
means of avalanches of evolutionary activity. This is appealing for a model
of biological evolution, since it has been observed that life on Earth could
be in a SOC state \cite{kauf,newm}. Nevertheless, models based in fitness
landscapes, or in a concept of fitness different from the biological one,
have been criticized from a biological point of view \cite{newm,jong}.


Since one of the characterizing aspects of life, and perhaps the most
fundamental one, is that of self-replication, it is our belief that more
realistic models should involve a dynamic population for each species. 
The starting point of combining population dynamics with evolution is the
association of the rates of birth and death and the carrying capacity with
phenotypes (observable features that arise from the genotype and are, then,
subject to mutation) \cite{roug}. The fitness, namely the expected number of
offsprings produced by an individual, arise from them. In this way, the
process of natural selection is directed by the ecological interactions
instead of by a non-biological notion of relative fitness.

Extinction is an essential component of evolution. The great majority of
species that have ever lived on Earth are now extinct\cite{raup}. 
There exist competing
hypothesis that account extinction as originating from within the biosystem,
or from external causes --what has been called ``bad genes or bad luck''. In
any case, the pattern of extinctions and of surviving species or groups of
species is certainly an interesting problem to model, to understand, and
eventually to check with the fossil record.

We show in this contribution a simple model of a large ecological
system in evolution. This produces features of extinction similar to those
claimed for the biosystem on Earth. We have chosen to study an ecological
model in which each species consists of a population interacting with the
others, that reproduces and evolves in time. The system is supposed to be a
food chain, and the interactions to be predator-prey. Mutations that change
the interactions are supposed to occur randomly at a low rate. Extinctions
of populations result from the predator-prey dynamics. This approach can be
thought as middle way between the microscopic simulation of ``artificial
life'' by Ray and others \cite{ray}, and the coarse-grain description of
models like BS's.

\section{The model}

Our model ecosystem consists of a number of species that interact and evolve
in time. In the course of its time evolution
the populations grow and shrink following a set of equations. 
Eventually, some of the species become
extinct as a result of their interaction with the others. Every now and then
we change one of the phenotypic features of one of the species,
mimicking a random mutation of its genome. This modification produces a
perturbation in the dynamics of the ecosystem, and eventually leads to the
extinctions.

To be more precise, let's consider a simple example of a food web, namely a
one-dimensional food chain. $N$ species interact in such a way that the
species $i$ feeds on the species $i-1$, and is eaten by the species $i+1$.
The species $1$ is an autotroph: it feeds at a constant rate on an
``environment''. The species $N$, the top of the chain, is not eaten by any
species, but dies giving its mass to the environment. Each species has a
population that evolves in time and interacts with its neighbors in the
chain. Furthermore, we consider this evolution in discrete time, which
is often more realistic than a continuous one \cite{roug} and simpler to
simulate in a computer.

As has been said above, each species acts as a predator with respect to the
one preceding it in the chain, and as a prey with respect to the one
following it. As a further simplification, we suppose that there are
no intrinsic birth and death rates, apart from those arising from the
predation and prey contributions. Let's propose the equations governing this
behaviors \cite{murray}. As a predator, the ``population'' (a continuous
density) of the species $i$, $n^i$, changes from time $t$ to time $t+1$
according to: 
\begin{equation}
\Delta n_t^i=k_in_t^{i-1}n_t^i\left( 1-n_t^i/c_i\right) ,  \label{predator}
\end{equation}
where $k_i$ is a rate of growth of the predator population and $c_i$ is a
carrying capacity that accounts for a limitation imposed by the environment.
Note that (\ref{predator}) includes this carrying capacity in a logistic
factor to avoid an unbounded growth of the population. Also observe that the
growth is proportional to the population of preys, without a ``satiation''
factor. Similarly, as a prey, the population of the species $i$ will
diminish according to: 
\begin{equation}
\Delta n_t^i=-g_in_t^in_t^{i+1}.
\end{equation}
The parameters $k_i$, $g_i$ and $c_i$ are the phenotypic features of our
species. In the course of the evolution we allow them to change,
mimicking random mutations. Moreover, they are the same for all the
individuals of each single population. We do not model races, traits,
polymorphisms or any phenotypic variation within a species, and when a
mutation occurs it is assumed that the whole population ``moves'' instantly
to the new state. In this sense, we are modelling the co-evolution of
the species and disregarding the evolution of a single one as well as
other important phenomena like the formation of new species \cite{vand}.

Combining the two roles of predator and prey that each species performs, and
the special status of the ends of the chain, we can write the following set
of equations for the evolution of the system: 
\begin{equation}
\left\{ 
\begin{array}{lcl}
\Delta n_t^i & = & k_in_t^{i-1}n_t^i\left( 1-n_t^i/c_i\right)
-g_in_t^{i+1}n_t^i \mbox{\ \ \ \ for $i=1 \dots N$} \\ 
n^0 & = & n^{N+1} = 1,
\end{array}
\right.  \label{model}
\end{equation}
where we have introduced two fictitious species, $0$ and $N+1$, to take
account for the border condition.

We make two simplifications to the system (\ref{model}): 1) we
assume that all the carrying capacities are equal, and equal to $1$;
2) we assume that $g_i=k_{i+1}$. In this way we reduce the number
of parameters that define the phenotypic features of the ecosystem.

The dynamics of the system is as follows. At time $t=0$, all the populations
and interactions are chosen at random with uniform distribution in the
interval $\left( 0,1\right) $. Then the populations begin to evolve
according to the system (\ref{model}). In the course of the evolution driven
by eq.(\ref{model}) a population can go to zero. As this can happen
asymptotically, we consider a species extinct if its population drops
below a given threshold. This is reasonable since actual biological populations
are discrete. In order to keep constant the number of species
 we replace an extinct one with a
new one, which can be thought as a species coming to occupy the niche left
by the extinct one \cite{footnote}. The new population, and the (two) new interactions with
its neighbors in the chain, are also drawn at random from a uniform
distribution in the interval $\left( 0,1\right) $. On top of this dynamics
of predation and extinctions, we introduce random mutations. In each time
step a mutation is produced with probability $p$; the species to mutate is
chosen at random and the mutation itself consists of the replacement of the
species with a new one, with a new population and new interactions with its
predators and preys (all random in $(0,1)$).

Observe that we do not introduce the fitness of a species as a dynamical
variable. We do not even need to compute it from the ``phenotypes'' $k_i$.
The fitness, the degree of adaptation of a species to the ecosystem, arises
from the phenotypes, the populations, and the dynamics, and it determines
whether a species will thrive or become extinct. Chance is introduced by
the random mutations (and the random replacement of extinct species). It
provides the material the natural selection works on. This, in turn,
determines the survival of the fittest by simply eliminating from the system
those species that cannot cope with the competing environment. We believe in
this way we avoid a fundamental problem in the models of evolution as a walk
in a fitness landscape, namely that the concepts of fitness is not the
biological one \cite{jong}.

\section{Results of the numerical simulation}

We have run our model for several chain sizes, ranging from $50$ to $1000$
sites, and for times of about $10^7$ steps. In the results reported below
we let the system evolve, during a transient period, from the initial
random state to an organized one.

In fig.\ \ref{pob} we show a typical evolution of the whole population, $%
\sum_1^N n^i$. 
Although each population greatly changes in the course of time (what
is not shown in the picture), we observe that the whole population
remains relatively stable. This is due to the saturation factor in the
predation term of the evolution equations.
This whole population shows a short time oscillatory dynamics
governed by the competition between species through the set of equations (%
\ref{model}), and a long time evolution characterized by periods of relative
stasis and periods of fast change. This feature is the effect of mutations
and extinction of some species. Without the extinctions and mutations, the
dynamics of the system should probably be chaotic. But it is not this
feature that we want to analyze here. Instead, we shall focus on the pattern
of extinctions.

As the set of $k_i$ represents the phenotypes of the whole ecosystem, its
distribution, $P(k)$, can be used to characterize its state. Let's observe what
happens in the course of time, including the transient mentioned above.
Initially the $k_i$ are chosen at random, and thus its distribution is flat
in $(0,1)$, with mean $0.5$. This is shown in fig.\ \ref{k} as a full line. As
time passes, and as a result of the dynamics, this distribution shifts to a
non-uniform one, as shown in fig.\ \ref{k} with dashed lines. 
The whole distribution shifts towards lower values of the interaction,
showing a tendency of the system to reduce the coupling between the species.
In the course of
the evolution this distribution fluctuates following the pattern of mutations
and extinctions, but preserves its form. 

Fig.\ \ref{kmed} shows the above mentioned fluctuations in $P(k)$
as the evolution of the mean value of $k_i$ in the
system, after the transient. 
It corresponds to the same run as fig.\ \ref{pob}, and the same time
window is shown. Similarly to that, it displays a pattern of periods of stasis
interrupted by periods of fast change, but without the short time
oscillations displayed by the population. 
There are periods of stasis of all lengths, to a degree that the unique
scale of the figure cannot display. This feature of a lack of a typical
length will be analyzed immediately. 
Observe in this figure that the mean value
oscillates around $0.24$, corresponding to a distribution like that
shown in fig.\ \ref{k} with a dotted line.

The extinction events also display this characteristic pattern of periods of
stasis and periods of change, without a typical size. 
In order to characterize this, we have chosen the time between 
two consecutive extinction events, $\tau$, which distribution is shown
in fig.\ \ref{ext} for several system sizes an 
probabilities of mutation. Observe that
they follow a power law for several decades of large values of $\tau$,
before a region where the effects of
the finite size of the system start to appear. This is a sign that the
system has self-organized into a critical state. 
In other words, the extinction events are
distributed in the time axis in such a way that the time between extinctions
does not have a characteristic duration --as should have if the 
distribution were exponential. Extinctions appear to come in
bursts, or avalanches, of any size.

In fig.\ \ref{aval} the pattern of extinction events of the system 
is seen in the course of time. 
The graphic displays time in the abscissa and the index in the food 
chain in the ordinate.
Each dot marks the moment in which a species has become
extinct. Each cross, a species that suffers a chance mutation. It can be
seen that some mutations trigger avalanches of extinction, and that these
propagate in the ``prey'' direction. (Bear in mind that an extinct species
is replaced by a random new one, most probably with a larger population than
its predecessor, and observe that this has a negative impact in the
corresponding {\em prey}.) It is also apparent that this avalanches have
a complex shape in space-time. It is not easy to measure their size since,
as can be seen in fig.\ \ref{aval}, they overlap. 
See, for example, a mutation that is {\em not} followed by any avalanche
(lower left), another that triggers a very small one (lower right),
and several that start events of varying size.
In any case, let's define a time step, $\Delta t$, divide the
time axis with it, and count the number of extinctions in each interval.
Now, let's call the fraction of species that have become extinct in each
interval the {\em size}, $S$, of the extinction. $S$ will obviously depend
on the time step and on the size of the system: $S=S(\Delta t,N)$. If the
system is in a critical state this function will obey some scaling law on
the variable $N$. In fig.\ \ref{scaling} we have scaled the distribution of
the system size $S(N)$, $P(S,N)$, obtained for different system sizes according
to the ansatz: 
\begin{equation}
P(S,N)=N^\beta f(S\cdot N^\nu ).
\end{equation}
We can observe that the four curves collapse to a single one, showing the
scaling behaviour that is typical of a critical state.

\section{Conclusions}
We have introduced a simple model of co-evolution and extinction 
in a food chain. This consists of a finite chain of species
of predators and preys. Their populations evolve in time following
Lotka-Volterra-like equations. Evolution is mimicked by randomly
changing a phenotype. Natural selection is provided by the deterministic
behaviour of the dynamical system, that produces the extinction
of any species that cannot cope with its interactions. No relative
fitness or fitness landscape had to be invoked. Nevertheless,
the pattern of extinctions displayed by this toy ecosystem appears to
be similar to that proposed for the biosystem on Earth.
Namely, the system seems to be in a critical state, in which
extinctions occur in avalanches. The time between extinctions, and the
lifetime of any species follow distributions that behave
like power-laws of time, implying that there is no typical size
for the time that a species remains in the system.
I should be of interest, in a future work, to study the precise
instability that produces the shift of the distribution of the
interactions towards low values. The analytical treatment of this
problem is currently under study.

The author greatly acknowledges Ruben Weht for invaluable discussions,
and thank Hilda Cerdeira for a careful reading of the manuscript.

\begin{figure}[tbp]
\caption{ Evolution of the whole population of a chain of $100$ species. 
Probability of mutation, 
$P_{mut}= 10^{-5}$. Observe the superposition of small oscillations 
and the much larger spikes signaling the mutations
and extinctions. }
\label{pob}
\end{figure}

\begin{figure}[tbp]
\caption{ Distribution, $P(k)$, of the interactions in a 
chain of 1000 species. Solid line: initial distribution; 
dashed line: $t=5000$; dotted line: $t=5\times10^{6}$. 
$P_{mut}=10^{-5}$. }
\label{k}
\end{figure}

\begin{figure}[tbp]
\caption{ Evolution of the mean value of the interactions, 
$\langle k \rangle
$, of a chain of $100$ species. $P_{mut}= 10^{-5}$.
The plot corresponds to the same run as that of fig.\ \ref{pob}.}
\label{kmed}
\end{figure}

\begin{figure}[tbp]
\caption{The distribution, $P(\tau)$, of time $\tau$ 
between two consecutive extinctions.
$P_{mut}= 10^{-5}$. }
\label{ext}
\end{figure}

\begin{figure}[tbp]
\caption{Space-time pattern of extinctions as they occur 
in a chain of $200$ species. The lower
plot shows a detail of the upper one. Each dot is an extinction 
event. Crosses, shown by arrows, indicate mutations. $P_{mut}=10^{-5}$.}
\label{aval}
\end{figure}

\begin{figure}[tbp]
\caption{Scaled distribution of the size of the extinctions, $P(S)$, as
a function of the scaled size, $S$. The four curves correspond to systems
of $N=50$, 100, 200 and 500 species. }
\label{scaling}
\end{figure}


\begin{references}
\bibitem{footnote}
If the number of species should not be kept constant, another assumption
should be made to somehow fill the hole in the chain so it doesn't break in
two with each extinction. For example, the prey and the predator of the
extinct species could become mutual prey and predator, with a new
interaction constant; but we think that the replacement of the niche with a
new species is biologically more relevant.
\end{references}
\end{document}